\documentclass[aps,prb,reprint,groupedaddress]{revtex4-1}
\usepackage{graphicx}
\begin{document}

%\preprint{}

\title{Local environment and oxidation state of Mn impurity in SrTiO$_3$
determined from XAFS data}


\author{I. A. Sluchinskaya}
\email[]{irinasluch@nm.ru}
%\thanks{}
\affiliation{Physics Department, Moscow State University, Moscow, 119991 Russia}
\author{A. I. Lebedev}
%\thanks{}
\affiliation{Physics Department, Moscow State University, Moscow, 119991 Russia}
\author{A. Erko}
%\thanks{}
\affiliation{BESSY II, Berlin, 12489 Germany}

\date{\today}

\begin{abstract}
The local environment and oxidation state of Mn impurity in strontium titanate
doped with 3\% Mn were studied by X-ray absorption fine structure spectroscopy.
The influence of the preparation conditions on the incorporation of the impurity
into the $A$ and $B$~sites was studied. It was established that Mn ions substituting
for Ti are in the Mn$^{4+}$ oxidation state and on-center. Mn ions substituting
for Sr are in the Mn$^{2+}$ oxidation state, are off-center and displaced from
the lattice sites by $\sim$0.32~{\AA}. It was demonstrated that studies of the
X-ray absorption near-edge structure can be used to determine the concentration
ratio of Mn atoms located at the $A$ and $B$~sites.

\texttt{DOI: 10.3103/S1062873810090145}
\end{abstract}

%\keywords{}

\maketitle


It has long been considered that upon doping of SrTiO$_3$ with manganese, the
impurity atoms substitute for Ti atoms and are in the Mn$^{4+}$ oxidation
state.~\cite{PhysRevLett.2.341}  Annealing of the samples in a reducing atmosphere
can convert the Mn ions to a lower oxidation state.~\cite{PhysRevB.16.4761}
Recently, Lemanov et al.~\cite{PhysSolidState.46.1442,JApplPhys.98.056102}
revealed the strong dielectric relaxations in SrTiO$_3$(Mn) at $T < 77$~K. They
explained the observed phenomena by reorientation of the dipoles of ``polaron''
defects and off-center Mn$^{4+}_{\rm Ti}$ ions; however, they did not associate
the phenomena with the appearance of ferroelectricity in strontium titanate.
Later, Tkach et al.~\cite{ApplPhysLett.86.172902} found the preparation
conditions in which the impurity Mn atoms can substitute for Sr atoms in the
Mn$^{2+}$ oxidation state. Unusual dielectric phenomena were observed in all
such samples. In contrast, in samples where the Mn atoms were at the Ti sites,
these effects were absent. In order to explain the dielectric properties of
Mn-doped samples, it was suggested that Mn$^{2+}$ ions are off-center when
substituting for the Sr atoms, and the observed relaxation phenomena are due
to the slowing down of their jumps between the potential minima when decreasing
temperature.~\cite{PhysRevB.73.104113} This supposition was confirmed by EPR
studies in Ref.~\onlinecite{PhysRevB.76.054104}.

The aim of this work was to determine the structural position and oxidation
state of the Mn impurity in SrTiO$_3$ samples prepared in various conditions
using X-ray absorption fine structure (XAFS) spectroscopy.

The samples studied in this work had nominal compositions of
(Sr$_{0.97}$Mn$_{0.03}$)TiO$_3$ and Sr(Ti$_{0.97}$Mn$_{0.03}$)O$_3$ and
were prepared by the solid-state synthesis method. The starting components were
SrCO$_3$, nanocrystalline TiO$_2$ obtained by hydrolysis of
tetrapropylortotitanate and dried at 500\,$^\circ$C, and
Mn(CH$_3$COO)$_2$$\cdot$4H$_2$O. The components were weighed in the required
proportions, ground in acetone until the mixture became completely dry, and
annealed in air at 1100\,$^\circ$C for 8~h. The obtained powders were ground
once again and annealed under the same conditions. Some of the samples were
additionally annealed in air at 1350\,$^\circ$C for 2.5~h and at 1500\,$^\circ$C
for 1~h. It should be noted that the use of a soluble manganese salt and its
impregnation into a mixture of SrCO$_3$ and mamocrystalline TiO$_2$ results
in a partial reaction of the salt with SrCO$_3$ and to the adsorption of
the manganese and strontium acetates on the TiO$_2$ nanoparticles with a large
surface area. Ultimately, this ensured a uniform distribution of the impurity
in the obtained samples.

Extended X-ray absorption fine structure (EXAFS) and X-ray absorption near-edge
structure (XANES) spectra were recorded in X-ray fluorescence mode on the KMC-2
station of the BESSY synchrotron radiation source in the vicinity of the
Mn $K$-edge (6.539~keV) at 300~K. EXAFS spectra were processed in a conventional
way. The details of the experiment can be found in Ref.~\onlinecite{JETPLett.89.457}.

\begin{figure}
\includegraphics[scale=1.1]{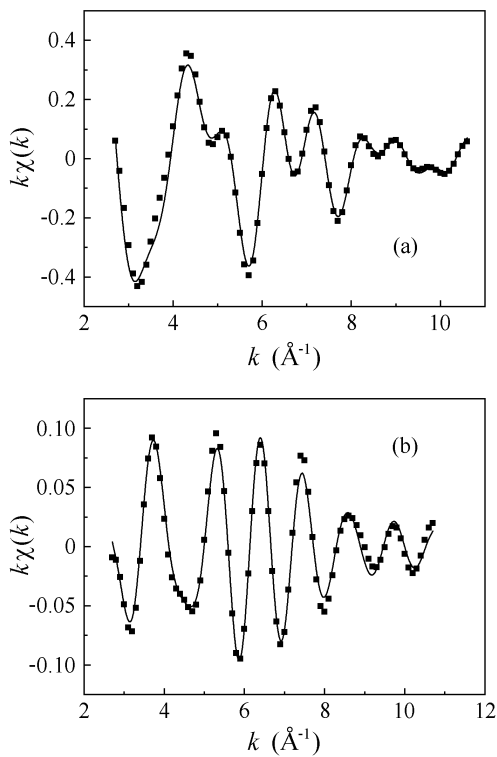}
\caption{EXAFS spectra of two SrTiO$_3$(Mn) samples recorded at the Mn $K$-edge:
(a) Sr(Ti$_{0.97}$Mn$_{0.03}$)O$_3$ sample annealed at 1100\,$^\circ$C; (b)
(Sr$_{0.97}$Mn$_{0.03}$)TiO$_3$ sample annealed at 1500\,$^\circ$C. The points
represent the experimental data; the lines are their best theoretical fits.}
\label{fig1}
\end{figure}

\begin{table*}
\caption{\label{table}Structural parameters obtained from the analysis of the
EXAFS spectra of two studied samples ($R_i$ is the distance to the $i$th shell,
and $\sigma^2_i$ is the Debye-Waller factor for this shell).}
\begin{ruledtabular}
\begin{tabular}{ccccc}
Sample & Shell & $R_i$ ({\AA}) & $\sigma^2_i$ ({\AA}$^2$) & Atom \\
\hline
Sr(Ti$_{0.97}$Mn$_{0.03}$)O$_3$ & 1 & 1.914        & 0.0012       & O \\
annealed at 1100\,$^\circ$C    & 2 & 3.328        & 0.0015       & Sr \\
                    & 3 & 3.895$^*$    & 0.0061$^*$   & Ti \\
(Sr$_{0.97}$Mn$_{0.03}$)TiO$_3$ & 1 & 2.32; 2.86   & 0.040; 0.021 & O \\
annealed at 1500\,$^\circ$C    & 2 & 3.095; 3.467 & 0.007; 0.008 & Ti \\
                    & 3 & 3.84         & 0.014        & Sr \\
\end{tabular}
\end{ruledtabular}
\noindent{\footnotesize $^*$The values for a single channel that takes into
account the focusing effect (double and triple scattering paths of multiple
scattering). \hfill}
\end{table*}

The EXAFS spectra for two samples, Sr(Ti$_{0.97}$Mn$_{0.03}$)O$_3$ annealed at
1100\,$^\circ$C and (Sr$_{0.97}$Mn$_{0.03}$)TiO$_3$ annealed at 1500\,$^\circ$C,
are shown in Fig.~\ref{fig1}. An analysis of the EXAFS spectra for the
Sr(Ti$_{0.97}$Mn$_{0.03}$)O$_3$ sample shows that the obtained spectra are
completely consistent with the model in which the Mn atoms substitute for the
Ti atoms and are on-center (see Table~\ref{table}). For the
(Sr$_{0.97}$Mn$_{0.03}$)TiO$_3$ sample, a good agreement between the experimental
and calculated EXAFS spectra (Fig.~\ref{fig1}(b)) can be obtained only in the
model with an off-center Mn atom displaced the from the Sr site, which
manifests in the appearance of two Mn--Ti distances (3.095 and 3.467~{\AA}, see
Table~\ref{table}). The displacement of the Mn atom from the $A$ site estimated
from the obtained Mn--Ti distances is $\sim$0.32~{\AA}. Our results on the
on-center position of the Mn$^{4+}$ ion at the Ti site and the off-center
position of the Mn$^{2+}$ ion at the Sr site agree with the results of the
theoretical calculations by Kvyatkovskii.~\cite{PhysSolidState.51.982}

\begin{figure}
\includegraphics{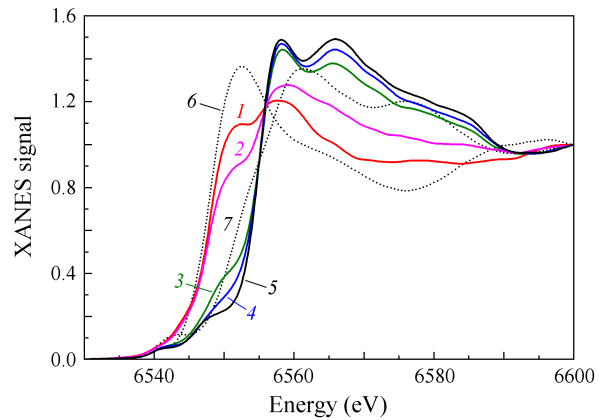}
\caption{(Color online) XANES spectra of five SrTiO$_3$(Mn) samples and two
reference compounds of divalent and tetravalent manganese: (1)
(Sr$_{0.97}$Mn$_{0.03}$)TiO$_3$
sample annealed at 1500\,$^\circ$C; (2) (Sr$_{0.97}$Mn$_{0.03}$)TiO$_3$ sample
annealed at 1350\,$^\circ$C; (3) (Sr$_{0.97}$Mn$_{0.03}$)TiO$_3$ sample annealed
at 1100\,$^\circ$C; (4) Sr(Ti$_{0.97}$Mn$_{0.03}$)O$_3$ sample annealed at
1100\,$^\circ$C; (5) Sr(Ti$_{0.97}$Mn$_{0.03}$)O$_3$ sample annealed at
1500\,$^\circ$C; (6) Mn(CH$_3$COO)$_2$$\cdot$4H$_2$O, (7) MnO$_2$.}
\label{fig2}
\end{figure}

The XANES spectra for five SrTiO$_3$(Mn) samples and two reference compounds
(Mn(CH$_3$COO)$_2$$\cdot$4H$_2$O and MnO$_2$) are shown in Fig.~\ref{fig2}.
A comparison of them shows that the absorption edges in the spectra of the
(Sr$_{0.97}$Mn$_{0.03}$)TiO$_3$ sample annealed at 1500\,$^\circ$C (curve~1)
and Sr(Ti$_{0.97}$Mn$_{0.03}$)O$_3$ sample annealed at 1100\,$^\circ$C (curve~4)
are shifted with respect to one another by $\sim$7~eV. It is widely
accepted that the oxidation state of an atom in a crystal can be determined
from the position of the steepest region on the absorption edge of the
atom in the XANES spectra, so the observed shift of the absorption edge
directly indicates that in the above two samples the Mn atoms, located
(according to the EXAFS data) at two different sites of the lattice, are
in two different oxidation states. By comparing our spectra with those of
the reference compounds, one may conclude that in SrTiO$_3$ the Mn ions located
at the $A$~sites are in +2 oxidation state, and the Mn ions located at the
$B$~sites are in +4 oxidation state. Some difference in the shape of the
absorption edge of the samples, in which the Mn ions are in +4 oxidation
state, and of the MnO$_2$ reference compound is due to the fact that
the position and the shape of the absorption edge in XANES spectra are
determined not only by the oxidation state of the atom, but also by the
band structure and crystal structure of the sample under investigation. At
room temperature, MnO$_2$ is a metal with the rutile structure, whereas the
studied samples are dielectrics with the cubic perovskite structure. The weak
pre-edge structure observed in all the samples is due to forbidden optical
transitions from the $1s$ level of the Mn atom to the conduction band (i.e.,
it is also determined by the band structure).

From a comparison of XANES spectra of the samples with a nominal composition
of (Sr$_{0.97}$Mn$_{0.03}$)TiO$_3$ annealed at 1350\,$^\circ$C (curve~2 in
Fig.~\ref{fig2}) and 1100\,$^\circ$C (curve~3 in Fig.~\ref{fig2}) and the
spectra of the samples for which the EXAFS analysis was performed (curves~1 and
4 in Fig.~\ref{fig2}), it follows that the spectra~2 and 3 can be regarded as
a superposition the of spectra~1 and 4. This means that in these samples, Mn is
at both lattice sites and in both oxidation states. From these data it follows
that in the samples with a deliberate deviation from Sr/Ti stoichiometry toward
the Sr deficit (the (Sr$_{0.97}$Mn$_{0.03}$)TiO$_3$ nominal composition), an
increase in the temperature of the final annealing results in a systematic
increase in the concentration of manganese in +2 oxidation state at the $A$~sites.
Indeed, in the sample annealed at 1100\,$^\circ$C (curve~3 in Fig.~\ref{fig2}),
the majority of Mn atoms are at the $B$~sites despite the deliberate deviation
of the sample composition from stoichiometry with the aim of incorporating
the impurity into the $A$~site. The thermal treatment of this sample at
1500\,$^\circ$C transfers the impurity Mn atoms from the Ti sites to the Sr
sites, in which their oxidation state is +2. As for the samples with a
deliberate deviation from stoichiometry toward the Ti defecit (the
Sr(Ti$_{0.97}$Mn$_{0.03}$)O$_3$ nominal composition), an increase in the
temperature of the final annealing results in an increase in the Mn concentration
at the $B$~sites in +4 oxidation state (curves~4 and 5).

In summary, the studies of XANES and EXAFS spectra of SrTiO$_3$(Mn) have shown
that, depending on the preparation conditions, Mn atoms can be incorporated into
the $A$ and $B$~sites of the perovskite structure and stay in them in different
oxidation states. The redistribution of the impurity atoms between two sites
can be controlled by the annealing temperature and a deliberate deviation of
the sample composition from stoichiometry. It was established that the impurity
Mn atoms substituting for Ti atoms occupy the on-center positions in the lattice
and are in +4 oxidation state, whereas the Mn atoms substituting for Sr atoms
are off-center, displaced from the lattice sites by $\sim$0.32~{\AA}, and
are in +2 oxidation state. Our direct data confirm the
supposition~\cite{ApplPhysLett.86.172902,PhysRevB.73.104113} about the possible
incorporation of the manganese impurity into the $A$~sites. It was demonstrated
that studies of the XANES structure can be used to determine the concentration
ratio of Mn atoms occupying different crystallographic positions in the lattice.

%\begin{acknowledgments}
%\end{acknowledgments}


%\bibliography{all}

\begin{thebibliography}{9}%
\makeatletter
\providecommand \@ifxundefined [1]{%
 \@ifx{#1\undefined}
}%
\providecommand \@ifnum [1]{%
 \ifnum #1\expandafter \@firstoftwo
 \else \expandafter \@secondoftwo
 \fi
}%
\providecommand \@ifx [1]{%
 \ifx #1\expandafter \@firstoftwo
 \else \expandafter \@secondoftwo
 \fi
}%
\providecommand \natexlab [1]{#1}%
\providecommand \enquote  [1]{``#1''}%
\providecommand \bibnamefont  [1]{#1}%
\providecommand \bibfnamefont [1]{#1}%
\providecommand \citenamefont [1]{#1}%
\providecommand \href@noop [0]{\@secondoftwo}%
\providecommand \href [0]{\begingroup \@sanitize@url \@href}%
\providecommand \@href[1]{\@@startlink{#1}\@@href}%
\providecommand \@@href[1]{\endgroup#1\@@endlink}%
\providecommand \@sanitize@url [0]{\catcode `\\12\catcode `\$12\catcode
  `\&12\catcode `\#12\catcode `\^12\catcode `\_12\catcode `\%12\relax}%
\providecommand \@@startlink[1]{}%
\providecommand \@@endlink[0]{}%
\providecommand \url  [0]{\begingroup\@sanitize@url \@url }%
\providecommand \@url [1]{\endgroup\@href {#1}{\urlprefix }}%
\providecommand \urlprefix  [0]{URL }%
\providecommand \Eprint [0]{\href }%
\providecommand \doibase [0]{http://dx.doi.org/}%
\providecommand \selectlanguage [0]{\@gobble}%
\providecommand \bibinfo  [0]{\@secondoftwo}%
\providecommand \bibfield  [0]{\@secondoftwo}%
\providecommand \translation [1]{[#1]}%
\providecommand \BibitemOpen [0]{}%
\providecommand \bibitemStop [0]{}%
\providecommand \bibitemNoStop [0]{.\EOS\space}%
\providecommand \EOS [0]{\spacefactor3000\relax}%
\providecommand \BibitemShut  [1]{\csname bibitem#1\endcsname}%
\let\auto@bib@innerbib\@empty
%</preamble>
\bibitem [{\citenamefont {M\"uller}(1959)}]{PhysRevLett.2.341}%
  \BibitemOpen
  \bibfield  {author} {\bibinfo {author} {\bibfnamefont {K.~A.}\ \bibnamefont
  {M\"uller}},\ }\href {\doibase 10.1103/PhysRevLett.2.341} {\bibfield
  {journal} {\bibinfo  {journal} {Phys. Rev. Lett.}\ }\textbf {\bibinfo
  {volume} {2}},\ \bibinfo {pages} {341} (\bibinfo {year} {1959})}\BibitemShut
  {NoStop}%
\bibitem [{\citenamefont {Serway}\ \emph {et~al.}(1977)\citenamefont {Serway},
  \citenamefont {Berlinger}, \citenamefont {M\"uller},\ and\ \citenamefont
  {Collins}}]{PhysRevB.16.4761}%
  \BibitemOpen
  \bibfield  {author} {\bibinfo {author} {\bibfnamefont {R.~A.}\ \bibnamefont
  {Serway}}, \bibinfo {author} {\bibfnamefont {W.}~\bibnamefont {Berlinger}},
  \bibinfo {author} {\bibfnamefont {K.~A.}\ \bibnamefont {M\"uller}}, \ and\
  \bibinfo {author} {\bibfnamefont {R.~W.}\ \bibnamefont {Collins}},\ }\href
  {\doibase 10.1103/PhysRevB.16.4761} {\bibfield  {journal} {\bibinfo
  {journal} {Phys. Rev. B}\ }\textbf {\bibinfo {volume} {16}},\ \bibinfo
  {pages} {4761} (\bibinfo {year} {1977})}\BibitemShut {NoStop}%
\bibitem [{\citenamefont {Lemanov}\ \emph {et~al.}(2004)\citenamefont
  {Lemanov}, \citenamefont {Smirnova}, \citenamefont {Sotnikov},\ and\
  \citenamefont {Weihnacht}}]{PhysSolidState.46.1442}%
  \BibitemOpen
  \bibfield  {author} {\bibinfo {author} {\bibfnamefont {V.~V.}\ \bibnamefont
  {Lemanov}}, \bibinfo {author} {\bibfnamefont {E.~P.}\ \bibnamefont
  {Smirnova}}, \bibinfo {author} {\bibfnamefont {A.~V.}\ \bibnamefont
  {Sotnikov}}, \ and\ \bibinfo {author} {\bibfnamefont {M.}~\bibnamefont
  {Weihnacht}},\ }\href {\doibase 10.1134/1.1788776} {\bibfield  {journal}
  {\bibinfo  {journal} {Phys. Solid State}\ }\textbf {\bibinfo {volume} {46}},\
  \bibinfo {pages} {1442} (\bibinfo {year} {2004})}\BibitemShut {NoStop}%
\bibitem [{\citenamefont {Lemanov}\ \emph {et~al.}(2005)\citenamefont
  {Lemanov}, \citenamefont {Sotnikov}, \citenamefont {Smirnova},\ and\
  \citenamefont {Weihnacht}}]{JApplPhys.98.056102}%
  \BibitemOpen
  \bibfield  {author} {\bibinfo {author} {\bibfnamefont {V.~V.}\ \bibnamefont
  {Lemanov}}, \bibinfo {author} {\bibfnamefont {A.~V.}\ \bibnamefont
  {Sotnikov}}, \bibinfo {author} {\bibfnamefont {E.~P.}\ \bibnamefont
  {Smirnova}}, \ and\ \bibinfo {author} {\bibfnamefont {M.}~\bibnamefont
  {Weihnacht}},\ }\href {\doibase 10.1063/1.2035313} {\bibfield  {journal}
  {\bibinfo  {journal} {J. Appl. Phys.}\ }\textbf {\bibinfo {volume} {98}},\
  \bibinfo {pages} {056102} (\bibinfo {year} {2005})}\BibitemShut {NoStop}%
\bibitem [{\citenamefont {Tkach}\ \emph {et~al.}(2005)\citenamefont {Tkach},
  \citenamefont {Vilarinho},\ and\ \citenamefont
  {Kholkin}}]{ApplPhysLett.86.172902}%
  \BibitemOpen
  \bibfield  {author} {\bibinfo {author} {\bibfnamefont {A.}~\bibnamefont
  {Tkach}}, \bibinfo {author} {\bibfnamefont {P.~M.}\ \bibnamefont
  {Vilarinho}}, \ and\ \bibinfo {author} {\bibfnamefont {A.~L.}\ \bibnamefont
  {Kholkin}},\ }\href {\doibase 10.1063/1.1920414} {\bibfield  {journal}
  {\bibinfo  {journal} {Appl. Phys. Lett.}\ }\textbf {\bibinfo {volume} {86}},\
  \bibinfo {pages} {172902} (\bibinfo {year} {2005})}\BibitemShut {NoStop}%
\bibitem [{\citenamefont {Tkach}\ \emph {et~al.}(2006)\citenamefont {Tkach},
  \citenamefont {Vilarinho}, \citenamefont {Kholkin}, \citenamefont {Pashkin},
  \citenamefont {Veljko},\ and\ \citenamefont {Petzelt}}]{PhysRevB.73.104113}%
  \BibitemOpen
  \bibfield  {author} {\bibinfo {author} {\bibfnamefont {A.}~\bibnamefont
  {Tkach}}, \bibinfo {author} {\bibfnamefont {P.~M.}\ \bibnamefont
  {Vilarinho}}, \bibinfo {author} {\bibfnamefont {A.~L.}\ \bibnamefont
  {Kholkin}}, \bibinfo {author} {\bibfnamefont {A.}~\bibnamefont {Pashkin}},
  \bibinfo {author} {\bibfnamefont {S.}~\bibnamefont {Veljko}}, \ and\ \bibinfo
  {author} {\bibfnamefont {J.}~\bibnamefont {Petzelt}},\ }\href {\doibase
  10.1103/PhysRevB.73.104113} {\bibfield  {journal} {\bibinfo  {journal} {Phys.
  Rev. B}\ }\textbf {\bibinfo {volume} {73}},\ \bibinfo {pages} {104113}
  (\bibinfo {year} {2006})}\BibitemShut {NoStop}%
\bibitem [{\citenamefont {Laguta}\ \emph {et~al.}(2007)\citenamefont {Laguta},
  \citenamefont {Kondakova}, \citenamefont {Bykov}, \citenamefont {Glinchuk},
  \citenamefont {Tkach}, \citenamefont {Vilarinho},\ and\ \citenamefont
  {Jastrabik}}]{PhysRevB.76.054104}%
  \BibitemOpen
  \bibfield  {author} {\bibinfo {author} {\bibfnamefont {V.~V.}\ \bibnamefont
  {Laguta}}, \bibinfo {author} {\bibfnamefont {I.~V.}\ \bibnamefont
  {Kondakova}}, \bibinfo {author} {\bibfnamefont {I.~P.}\ \bibnamefont
  {Bykov}}, \bibinfo {author} {\bibfnamefont {M.~D.}\ \bibnamefont {Glinchuk}},
  \bibinfo {author} {\bibfnamefont {A.}~\bibnamefont {Tkach}}, \bibinfo
  {author} {\bibfnamefont {P.~M.}\ \bibnamefont {Vilarinho}}, \ and\ \bibinfo
  {author} {\bibfnamefont {L.}~\bibnamefont {Jastrabik}},\ }\href {\doibase
  10.1103/PhysRevB.76.054104} {\bibfield  {journal} {\bibinfo  {journal} {Phys.
  Rev. B}\ }\textbf {\bibinfo {volume} {76}},\ \bibinfo {pages} {054104}
  (\bibinfo {year} {2007})}\BibitemShut {NoStop}%
\bibitem [{\citenamefont {Lebedev}\ \emph {et~al.}(2009)\citenamefont
  {Lebedev}, \citenamefont {Sluchinskaya}, \citenamefont {Erko},\ and\
  \citenamefont {Kozlovskii}}]{JETPLett.89.457}%
  \BibitemOpen
  \bibfield  {author} {\bibinfo {author} {\bibfnamefont {A.~I.}\ \bibnamefont
  {Lebedev}}, \bibinfo {author} {\bibfnamefont {I.~A.}\ \bibnamefont
  {Sluchinskaya}}, \bibinfo {author} {\bibfnamefont {A.}~\bibnamefont {Erko}},
  \ and\ \bibinfo {author} {\bibfnamefont {V.~F.}\ \bibnamefont {Kozlovskii}},\
  }\href {\doibase 10.1134/S0021364009090070} {\bibfield  {journal} {\bibinfo
  {journal} {JETP Lett.}\ }\textbf {\bibinfo {volume} {89}},\ \bibinfo {pages}
  {457} (\bibinfo {year} {2009})}\BibitemShut {NoStop}%
\bibitem [{\citenamefont {Kvyatkovskii}(2009)}]{PhysSolidState.51.982}%
  \BibitemOpen
  \bibfield  {author} {\bibinfo {author} {\bibfnamefont {O.~E.}\ \bibnamefont
  {Kvyatkovskii}},\ }\href {\doibase 10.1134/S1063783409050163} {\bibfield
  {journal} {\bibinfo  {journal} {Phys. Solid State}\ }\textbf {\bibinfo
  {volume} {51}},\ \bibinfo {pages} {982} (\bibinfo {year} {2009})}\BibitemShut
  {NoStop}%
\end{thebibliography}
\providecommand{\BIBYu}{Yu}

\end{document}